\documentclass[preprint,aps]{revtex4}

\usepackage{graphicx}% Include figure files
\usepackage{dcolumn}% Align table columns on decimal point
\usepackage{bm}% bold math

\begin{document}

%\markboth{Xiao-Gang He, A. Zee} {Geometric Mean Neutrino Mass
%Relation}

%%%%%%%%%%%%%%%%%%%%% Publisher's Area please ignore %%%%%%%%%%%%%%
%\catchline{}{}{}{}{}
%%%%%%%%%%%%%%%%%%%%%%%%%%%%%%%%%%%%%%%%%%%%%%%%%%%%%%%%%%%%%%%%%%%

\begin{center}
{\Large {\bf Geometric Mean Neutrino Mass Relation}}
\end{center}

\begin{center}
{Xiao-Gang He$^{1}$, and A. Zee$^2$}\\
\vspace*{0.3cm}

$^1$Department of Physics and Center for Theoretical Sciences,
National Taiwan University, Taipei, Taiwan\\

$^2$Kavli Institute for Theoretical Physics, University of
California Santa Barbara, CA 93106 USA
\end{center}
%\maketitle

%\pub{Received (Day Month Year)}{}%Revised (Day Month Year)}

%\begin{abstract}
\begin{center}
\begin{minipage}{12cm}

\noindent Abstract:\\
Present experimental data from neutrino oscillations have provided
much information about the neutrino mixing angles. Since neutrino
oscillations only determine the mass squared differences $\Delta
m^2_{ij} = m^2_i - m^2_j$, the absolute values for neutrino masses
$m_i$ can not be determined using data just from oscillations. In
this work we study implications on neutrino masses from a
geometric mean mass relation $m_2=\sqrt{m_1 m_3}$ which enables
one to determined the absolute masses of the neutrinos. We find
that the central values of the three neutrino masses and their
$2\sigma$ errors to be $m_1 = (1.58\pm 0.18)\mbox{meV}$, $m_2 =
(9.04\pm 0.42)\mbox{meV}$, and $m_3 = (51.8\pm 3.5)\mbox{meV}$.
Implications for cosmological observation, beta decay and
neutrinoless double beta decays are discussed.
%A mass matrix which
%produces the geometric mean mass relation is suggested.

\keywords{neutrino; mass; geometric.}
%\end{abstract}
\end{minipage}

\end{center}
%\ccode{PACS Nos.: include PACS Nos.}

There are abundant data\cite{pdg,data} from solar, atmospheric,
laboratory and long baseline neutrino experiments on neutrino mass
and mixing. Neutrino oscillations provide direct evidence of
non-zero neutrino masses and mixing between different species of
neutrinos. The mixing can be well described by the
Pontecorvo-Maki-Nakagawa-Sakata (PMNS) mixing matrix\cite{pmns} in
the weak interaction with three neutrino oscillations. The PMNS
can be parameterized as
\begin{eqnarray}
V = \left ( \begin{array}{lll}
c_{12}c_{13}&s_{12}c_{13}&s_{13}e^{-i\delta_{13}}\\
-s_{12}c_{23}-c_{12}s_{23}s_{13}e^{i\delta_{13}}&
c_{12}c_{23}-s_{12}s_{23}s_{13}e^{i\delta_{13}}&s_{23}c_{13}\\
s_{12}s_{23}-c_{12}c_{23}s_{13}e^{i\delta_{13}}&-c_{12}s_{23}
-s_{12}c_{23}s_{13}e^{i\delta_{13}}&c_{23}c_{13}
\end{array} \right )P,\label{mixing}
\end{eqnarray}
where $s_{ij} = \sin\theta_{ij}$, $c_{ij} = \cos\theta_{ij}$. For
Dirac neutrinos $P$ is a unit matrix, and for Majorana neutrinos
$P$ is a diagonal phase matrix with two independent phases and can
be written as $P= diag(1,exp[i\phi_2], exp[i\phi_3])$.

Neutrino oscillations also depend on the mass squared differences
$\Delta m^2_{ij} = m^2_i - m^2_j$ of neutrino masses $m_i$. The
present experimental information on the mixing angles and the mass
squared differences $\Delta m^2_{ij}$ can be summarized as the
following\cite{pdg,data}
\begin{eqnarray}
&&\sin^2\theta_{12} =
0.314(1^{+0.18}_{-0.15}),\;\;\sin^2\theta_{23} =
0.45(1^{+0.35}_{-0.20}),\;\;\sin^2\theta_{13} =
(0.8^{+2.3}_{-0.8})\times 10^{-2},\nonumber\\
&&\Delta m^2_{21} = 7.92(1\pm 0.09)\times
10^{-5}\mbox{eV}^2,\;\;\vert\Delta m^2_{32}\vert =
2.6(1^{+0.14}_{-0.15})\times 10^{-3}\mbox{eV}^2.
\end{eqnarray}
The errors are at $2\sigma$ level. No CP violating experiments in
neutrino oscillation have been performed, the phase $\delta_{13}$
is not known. Neutrino oscillations do not provide any information
about the Majorana phases.

Although there are stringent constraints on neutrino masses from
laboratory and cosmological data, and precise measurements of mass
squared differences from neutrino oscillations, the absolute
masses are not known. There are mechanisms proposed to understand
the smallness of the neutrino masses, such as
see-saw\cite{see-saw} and radiative loop generation of
masses\cite{zee}, a definitive mechanism to determine the absolute
values of neutrino masses is still lacking. It is desirable to
find some additional information, experimental or theoretical, to
determine the masses. Attempts using various ansatzes have been
made previously\cite{ansatz}. Here we find another interesting
relation which can lead to the determination of neutrino masses.
This is the geometric mean mass relation\cite{xing} $m_2
=\sqrt{m_1 m_3}$. We have chosen to work in the basis where the
values of neutrino masses are all positively defined.

Geometric mean mass relation has been considered for quarks
previously, in particular was used to predict the top quark
mass\cite{he}. To have some ideas whether this is a reasonable
attempt to pursue, in Fig.1 we summarize the values for
$log_{10}(m_i)$ for quarks and charged leptons. In the figure we
have used the central values for the quark masses at $\mu = m_Z$
with\cite{koide}: $m_u = (2.22\pm 0.24^{+0.14}_{-0.17})$ MeV, $m_d
= (4.42\pm 0.29^{+0.29}_{-0.34})$ MeV, $m_s = (84.7\pm
7.2^{+5.5}_{-6.6})$ MeV, $m_c = (0.661\pm
0.012^{+0.042}_{-0.047})$ GeV, $m_b = (2.996\pm
0.036^{+0.069}_{-0.074})$ GeV and $m_t = (180 \pm 13\pm 0.02)$
GeV.

Using $log(m_i/eV)$ as vertical axis, the geometric mean mass
relation is represented by a straight line for equal horizontal
interval of an arbitrary unit since $(log(m_2/eV)-log(m_1/eV))$/$
(log(m_3/eV)- log(m_2/eV))$ = $log(m_2/m_1)/log(m_3/m_2)$ =$ 1$.
The various $log$ plots are shown in Figure 1. Numerically we have
$log_{10} (m_c/m_u)/log_{10} (m_t/m_c)$ =$ 1.016(1\pm 0.166)$,
$log_{10}( m_s/m_d)/log_{10} (m_b /m_s)$ =$
 0.828(1\pm0.327)$, and
$log_{10} (m_\mu / m_e)/log_{10} (m_\tau / m_\mu)$ = $1.889$. We
see that the geometric mean mass relation holds well for the up
quarks, u, c and t. For the down quarks, d, s and b, the relation
holds at one $\sigma$ level. Unfortunately the best measured
charged lepton masses obviously do not satisfy the geometric mean
mass relation. The electron mass seems to be anomalously small for
some unknown reason. Alternatively, the tau mass is anomalously
low and/or the muon mass is anomalously high. One is tempted to
speculate that there is a mechanism that drives the electron mass
to zero.

\begin{figure}[!htb]
\begin{center}
\begin{tabular}{cc}
\includegraphics[width=7cm]{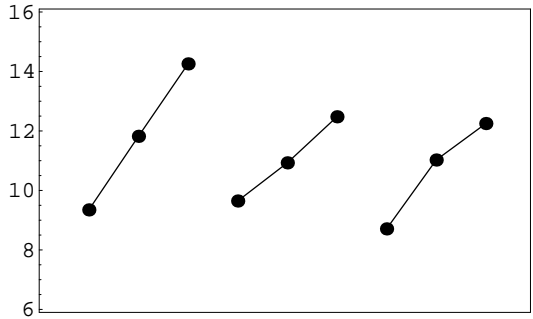}
\end{tabular}
\end{center}
\label{figure} \caption{Summary of central values for
$log_{10}(m_i/eV)$ with equal intervals of horizontal axis for
each point. From left to right, the lines are for up-quarks, down
quarks and charged leptons.} \label{diagram}
\end{figure}

With the geometric mass relation $m_2 = \sqrt{m_1 m_3}$, we have
\begin{eqnarray}
m_1^2 = {\Delta m^2_{21} \Delta m^2_{21}\over \Delta m^2_{32} -
\Delta m^2_{21}},\;\; m_2^2 = {\Delta m^2_{21} \Delta
m^2_{32}\over \Delta m^2_{32} - \Delta m^2_{21}},\;\; m_3^2 =
{\Delta m^2_{32} \Delta m^2_{32}\over \Delta m^2_{32} - \Delta
m^2_{21}}.
\end{eqnarray}
Experimental data from neutrino oscillation on mass squared
differences then determine the central values and $2\sigma$ errors
for the neutrino masses to be
\begin{eqnarray}
&&m_1 = (1.58\pm 0.18)\mbox{meV},\nonumber\\
&&m_2 = (9.04\pm 0.42)\mbox{meV},\nonumber\\
&&m_3 = (51.8\pm 3.5)\mbox{meV}.
\end{eqnarray}

The masses obtained above must be checked against known
experimental constraints. One of the most stringent constraints
comes from cosmology consideration. In contrast to oscillation
experiments, the contribution of the neutrinos to the energy
density of the universe, $\Omega_\nu h^2 \approx m_{sum}
/(93.5\;\mbox{eV})$ depends on the values of $m_{sum} =
m_1+m_2+m_3$ of course. The power spectrum of density perturbation
also depends on $m_{sum}$. The present bound obtained from
combining available data\cite{data} from CMB, large scale
structure power spectrum, baryonic acoustic oscillation and small
scale primordial spectrum from Lyman-alpha forest clouds, is very
stringent with $m_{sum} < 170$ meV. Using the geometric mean mass
relation, we obtain
\begin{eqnarray}
m_{sum} = (62.4\pm 3.5)\mbox{meV}.
\end{eqnarray}

It is interesting to note that this value is just 3 times smaller
than the current bound from cosmology and may be probed in the
future. In the near future $m_{sum}$ can be probed down to 120 meV
by Planck experiment which is still about two times above the
predicted value. However when combined with other data, the
predicted range can be probed. For example sky survey with an
order of magnitude larger survey volume would allow the
sensitivity to reach 30 meV\cite{weiler}. The mass ranges
predicted with the geometric mean mass relation may be tested in
the future.

There are other experimental constraints on neutrino masses. Of
particular interests to neutrino masses are constraints on
effective masses $\langle m_{\beta}\rangle$ and $m_{\beta \beta}$
from tritium $\beta$ decay and neutrinoless double $\beta$ decay,
respectively. $m_\beta$ has not been measured and the present
$2\sigma$ level upper bound, combing the Mainz and Triotsk
experiments, is $m_\beta < 1.8$ eV\cite{data}.  Planned
experiments, KATRIN and MARE, can reach a sensitivity about 0.2
eV\cite{dataf1}. Currently it is still in debating whether a
non-zero $m_{\beta\beta}$ has been measured in neutrinoless double
beta decay of $^{76}Ge$. If the claimed observation is true it
would imply a $2\sigma$ range\cite{data,klap} $0.43 \mbox{eV} <
m_{\beta\beta} < 0.81\mbox{eV}$. Future experiments can reach a
sensitivity as low as 9 meV\cite{dataf2}.

We now discuss the implications of the masses obtained from
geometric mean mass relation on $\langle m_{\beta}\rangle$ and
$m_{\beta\beta}$. These quantities are defined as
\begin{eqnarray}
&&\langle m_{\beta}\rangle = (m_{1}^2 |V_{e1}|^2 + m_{2}^2
|V_{e2}|^2 + m_{3}^2 |V_{e3}|^2)^{1/2}=(m_{1}^2 c^2_{12}c_{13}^2 +
m_{2}^2 s^2_{12}c^2_{13} +
m_{3}^2 s_{13}^2)^{1/2},\nonumber\\
 && m_{\beta\beta}| = |m_1 V_{e1}^2 + m_{2} V_{e2}^2 + m_{3}
V_{e3}^2|=|m_1 c^2_{12}c^2_{13}+m_2 s^2_{12}c^2_{13}e^{2i\phi_2} +
m_3 s^2_{13}e^{2i\phi_3}|. \label{double}
\end{eqnarray}

To have detailed information on $\langle m_{\beta}\rangle$ and
$m_{\beta \beta}$ one needs to have more information on the mixing
angles. A popular mixing matrix consistent with data is the
so-called tri-bimaximal mixing\cite{tribi} where $(V_{e1}, V_{e2},
V_{e3}) = (2/\sqrt{6}, 1/\sqrt{3},0)$. In this case the values for
$m_i$ predicted by the geometric mass relation would give a range
$(5.37\pm0.24)$ meV for $\langle m_{\beta}\rangle$. The range for
$m_{\beta\beta}$ depends on the unknown Majorana phase $\phi_2$.
However since the term proportional to $m_2$ again dominates, the
effect of $\phi_2$ is small. The value for $m_{\beta\beta}$ is
$\langle m_\beta \rangle/\sqrt{3}$ to a good approximation.

Finally we discuss how mass matrix which generating the geometric
mean mass relation can be constructed. To this end we note a
simple mass matrix of the form
\begin{eqnarray}
M_\nu = \left ( \begin{array}{lll} 0&0&a\\0&a&0\\a&0&b
\end{array}\right ),
\end{eqnarray}
with $b>>a$ gives,
\begin{eqnarray}
m_1 \approx - {a^2\over b},\;\;m_2 = a,\;\;m_3 \approx b +
{a^2\over b}.
\end{eqnarray}
The minus sign for $m_1$ can be removed by a redefinition of the
phases of neutrino fields. The above satisfies the geometric mean
mass related to a good approximation for neutrinos since $b$ is
about 6 times larger than $a$, $a/b <<1$.

It is however a challenge to have a model which naturally give the
above mass matrix. It is not difficult to have the texture zeros
in the above matrix. For example, if there is a $Z_8$ discrete
symmetry with the elements $Exp[i 2 n\pi/8]$ acts on leptons with
two Higgs doublets, and the quantum numbers of the leptons and the
two Higgs doublets are: $n$ for the left-haded lepton doublet
$l_n$ and the right-handed charged lepton $e_{Rn}$, and  ``0'' and
``2'' for $H_{0,2}$, the dimension-5 Weinberg operator
$\lambda_{ij} \bar l_i^c l_j H_k H_l$ will generate a mass matrix
of the form
\begin{eqnarray}
M_\nu = \left ( \begin{array}{lll}
0&0&a_{13}\\0&a_{22}&0\\a_{13}&0&a_{33}
\end{array}\right ).
\end{eqnarray}
The above admits the desired form if $a_{22} = a_{13}$. Of course
this amounts to the proposed geometric relation. We have not been
able to derive $a_{22} = a_{13}$ from some symmetry principles.
Further investigation is needed.

To summarize, we have studied the consequences of neutrino masses
with the geometric mean relation $m_2 =\sqrt{m_1m_3}$. With this
condition the neutrino masses can be determined from measured
mass-squared differences from oscillation experiments. We find
that the neutrino masses are $m_1 = (1.58\pm 0.18)\mbox{meV}$,
$m_2 = (9.04\pm 0.42)\mbox{meV}$, and $m_3 = (51.8\pm
3.5)\mbox{meV}$. Although these masses are small, they can be
probed by experiments from CMB measurements and large scale
structure survey. We have suggested a mass matrix which produces
the geometric mean mass relation, but we have not been able to
derive it from some symmetry principles. It is interesting to see
if a complete model with the geometric mean mass relation for
neutrino masses can be constructed.
\\

\noindent{\bf Acknowledgments} This work was supported in part by
NSC, and by NSF under grants PHY99-07949 and PHY00-98395 of USA.
XGH is also partially supported by NCTS. AZ is grateful to the
Radclife Institute for Advanced Study, Harvard University, and the
National Taiwan University where this work was done, for their
warm hospitality.

%\tighten

\end{document}